\documentclass[%
reprint,
superscriptaddress,
 amsmath,amssymb,
 aip,cha
]{revtex4-1}

\usepackage{amsmath}
\usepackage{subfig}
\usepackage{blindtext}
\usepackage{graphicx}% Include figure files
\usepackage{dcolumn}% Align table columns on decimal point
\usepackage{bm}% bold math
\usepackage{color}
\begin{document}

\preprint{APS/123-QED}

\title{Experimental observation of chimera and cluster states in a minimal globally coupled network}% Force line breaks with \\

\author{Joseph D. Hart}
 \affiliation{Institute for Research in Electronics and Applied Physics, University of Maryland, College Park, Maryland 20742, USA}
 \affiliation{Department of Physics, University of Maryland, College Park, Maryland 20742, USA}
\author{Kanika Bansal}%
\noaffiliation

\author{Thomas E. Murphy}
 \affiliation{Institute for Research in Electronics and Applied Physics, University of Maryland, College Park, Maryland 20742, USA}
\affiliation{Department of Electrical and Computer Engineering, University of Maryland, College Park, Maryland 20742, USA}

\author{Rajarshi Roy}
\affiliation{Institute for Research in Electronics and Applied Physics, University of Maryland, College Park, Maryland 20742, USA}
 \affiliation{Department of Physics, University of Maryland, College Park, Maryland 20742, USA}
 \affiliation{Institute for Physical Science and Technology, University of Maryland, College Park, Maryland, 20742, USA}

\date{\today}% It is always \today, today,

\begin{abstract}
A ``chimera state'' is a dynamical pattern that occurs in a network of coupled identical oscillators when the symmetry of the oscillator population is broken into synchronous and asynchronous parts. We report the experimental observation of chimera and cluster states in a network of four globally coupled chaotic opto-electronic oscillators. This is the minimal network that can support chimera states, and our study provides new insight into the fundamental mechanisms underlying their formation. We use a unified approach to determine the stability of all the observed partially synchronous patterns, highlighting the close relationship between chimera and cluster states as belonging to the broader phenomenon of partial synchronization. Our approach is general in terms of network size and connectivity. We also find that chimera states often appear in regions of multistability between global, cluster, and desynchronized states.

\end{abstract}

\maketitle
% * <jdhart12@gmail.com> 2015-12-14T23:14:32.771Z:
\begin{quotation}
We provide experimental evidence of chimera and cluster synchronous states in a globally coupled network of four opto-electronic oscillators. Since this is the minimal network in which a chimera state can occur, our apparatus provides the ability to experimentally test some of the fundamental properties of chimera states. Cluster synchronization has thus far been studied independently of chimera states; however, here we present a unified approach that exploits the symmetries in the network to determine the stability of chimeras and clusters. We obtain two important results: A) we provide a first experimental demonstration that chimeras can appear in small networks, contrary to the conventional assumption that a large network with non-local coupling is necessary \cite{panaggio2015chimera}, and B) we show that both cluster states and chimera states can be regarded as special cases of the more general phenomenon of partial synchronization.  The methods apply to networks of different size and topology, opening up potential applications to chimeras and other partial synchrony patterns in real world networks such as power grids.

\end{quotation}
\section{Introduction}

Since their original discovery \cite{kuramoto2002coexistence,abrams2004chimera}, there has been a great deal of discussion about the definition of chimera states and the conditions for their existence. It was originally thought that chimeras could exist only in large networks of non-locally coupled oscillators and only from special initial conditions \cite{panaggio2015chimera}. These assumptions were reflected in the decade-long gap between their theoretical discovery\cite{kuramoto2002coexistence} and the first experimental realization of chimeras in a spatial light modulator feedback system \cite{hagerstrom2012experimental} and chemical oscillator system \cite{tinsley2012chimera}. However, recent studies have shown that chimeras can actually appear in a much wider variety of networks: chimeras have now been observed experimentally in a mechanical system of metronomes \cite{martens2013chimera}, optical frequency combs \cite{viktorov2014coherence}, electrochemical systems \cite{schmidt2014coexistence}, star networks of Lorenz oscillators \cite{sinha2015chimera}, and electronic and opto-electronic delay systems \cite{larger2013virtual,rosin2014transient,larger2015laser}. This suggests that chimeras may exist more widely than at first expected. Indeed, recent theoretical work has found chimeras in small networks \cite{ashwin2015weak,panaggio2015chimera}, from random initial conditions \cite{daido2006diffusion}, and for global coupling \cite{kaneko1990clustering,daido2006diffusion,yeldesbay2014chimeralike,sethia2014chimera,schmidt2015clustering,schmidt2015chimeras}--well beyond the conditions initially assumed necessary for their existence.

While chimeras can exist in many different systems, one common characteristic seems to be that chimeras often appear in regions of multistability with other synchronous patterns \cite{kaneko1990clustering,larger2015laser,chandrasekar2014mechanism,martens2013chimera}. Recently, B\"ohm et al. proposed a network of four globally coupled lasers in which chimera states can emerge from random initial conditions and have linked the emergence of chimeras to a multistable region of parameter space \cite{bohm2015amplitude}.

Chimeras are a special type of partially synchronous state. The study of cluster synchrony, another type of partially synchronous state, has developed independently of chimeras; however, recent work has begun to link the existence of chimeras in globally coupled networks to clusters \cite{schmidt2015clustering,schmidt2015chimeras}. One major development in the study of cluster synchrony has been the ability to determine the clusters that are allowed to form from the symmetries in the network topology and to exploit those symmetries to derive the variational equations for stability calculations \cite{pecora2014cluster,sorrentino2015complete}. An extension of this theory has recently been used to explain phase-lag synchronization in networks of chemical oscillators \cite{totz2015phase}; however, as far as we are aware, this powerful new method has not yet been applied to study the existence or stability of chimera states.   

In this paper, we report on the experimental observation of chimera and other partially synchronous states in a minimal network of four identical, globally coupled opto-electronic oscillators with time-delayed feedback and coupling. We show that these states emerge from partial (or subgroup) symmetries in the network topology, and we calculate their linear stability using methods \cite{pecora2014cluster,sorrentino2015complete} recently developed for the study of cluster synchronization, highlighting that chimera and cluster states are closely related patterns of partial synchrony. We conclude with a discussion of the importance of multistability of partially synchronous states for the existence of chimera states.

\section{Partial synchronization in an opto-electronic network}
\subsection{Experimental apparatus}
\begin{figure}
\includegraphics[width=0.5\textwidth]{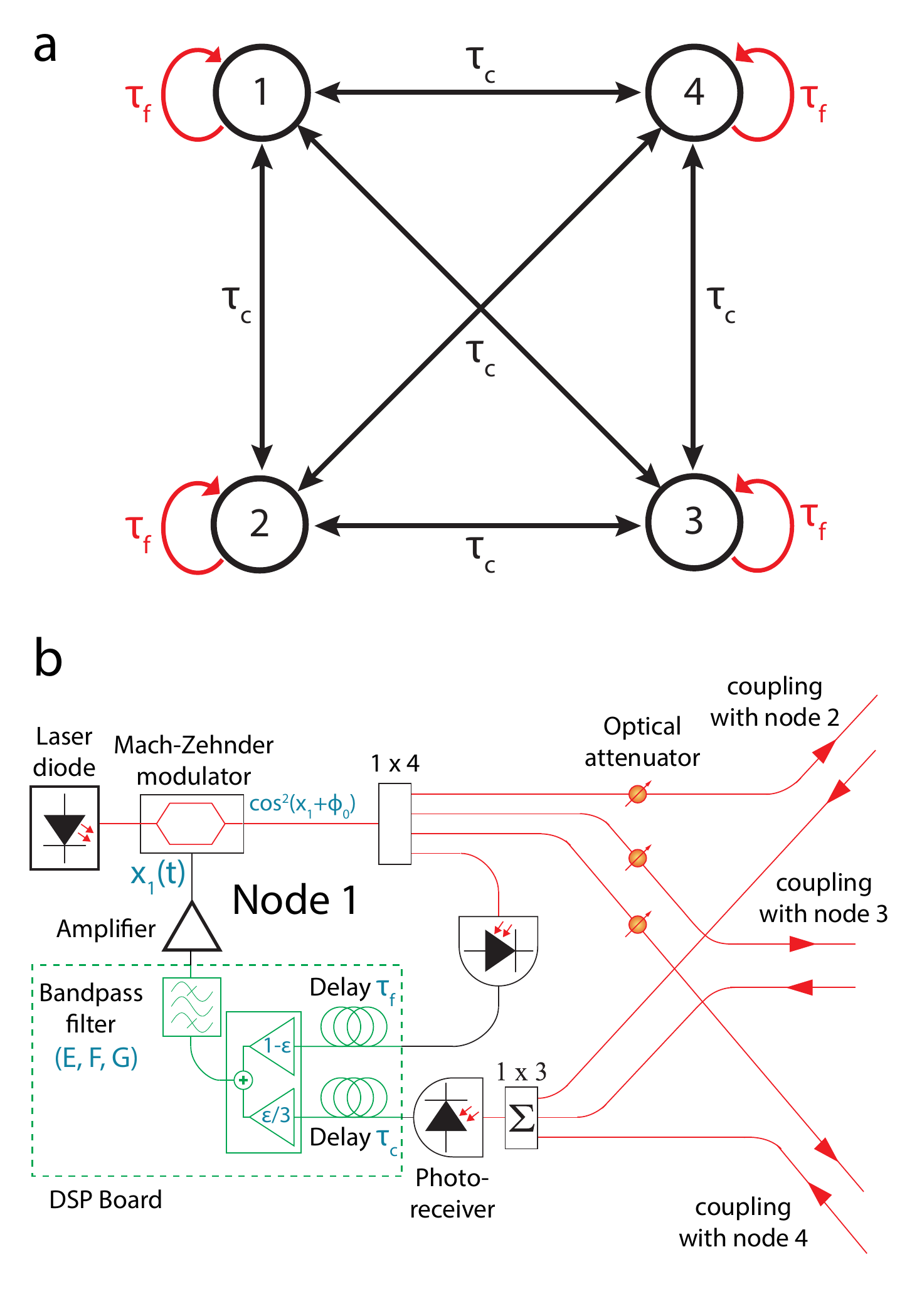}
\caption{\label{fig:exp} (a) Sketch of our globally coupled network. Each node has a self-feedback with feedback time delay $\tau_f$ (red) and is coupled to every other node with coupling time delay $\tau_c$. (b) Experimental schematic of a single node, showing the coupling to the other nodes. Optical connections are shown in red, electronic in black.}
\end{figure}

The experiment consists of a network of four globally coupled identical, opto-electronic, time-delayed feedback loops, whose individual and coupled dynamics have been studied previously \cite{murphy2010complex,ravoori2011robustness,williams2013experimental,2013synchronization, hart2015adding, larger2013complexity}. A layout of the network and a schematic of a single node are shown in Fig. \ref{fig:exp}. As shown in Fig. 1b, each node consists of a fiber-coupled laser diode whose light passes through a Mach-Zehnder modulator (MZM) with $V_\pi=3.4 $ V and is converted to an electrical signal by a photoreceiver. This electrical signal is delayed and filtered by a digital signal processing (DSP) board (Texas Instruments TMS320C6713) before being amplified and fed back to drive the MZM. The normalized voltage $x(t)\equiv\frac{\pi v_i(t)}{2V_\pi}$ applied to the MZM is measured and used as our dynamical variable. The digital filter is a two-pole Butterworth bandpass filter with cutoff frequencies $\omega_H/2\pi=100$ Hz and $\omega_L/2\pi=2.5$ kHz and a sampling rate of 24 kSamples/s. The output of the DSP is amplified such that each feedback loop has identical normalized round-trip gain $\beta=3.8$. Each node also has the same feedback time delay $\tau_f=1.4$ ms. We choose the phase bias $\phi_0=\pi/4$ so that for large $\beta$ the MZM nonlinearity becomes important and an uncoupled oscillator behaves chaotically.
 We choose the parameters such that a single uncoupled node behaves chaotically.
 
The incoming coupling signals are optically combined and converted into a second electrical signal by a second photodiode. The DSP board receives this second electrical signal, implements the filtering and coupling delay $\tau_c$ (which is in general different from the feedback delay $\tau_f$), and couples this second electrical signal with the first (feedback) signal.

The equations governing the dynamics of the network of opto-electronic oscillators are derived in ref. \cite{murphy2010complex} and are given by

\begin{equation}\label{eq:ueq}
\dot{\mathbf{u}}_i(t)=\mathbf{E}\mathbf{u}_i(t)-\mathbf{F}\beta\cos^2 (x_i(t)+\phi_0),
\end{equation}

\begin{equation}\label{xeq}
x_i(t)=\mathbf{G}\Big(\mathbf{u}_i(t-\tau_f)\\+\frac{\varepsilon}{n_{in}}\sum_{j}A_{ij}\big(\mathbf{u}_j(t-\tau_c)-\mathbf{u}_i(t-\tau_f)\big)\Big)
\end{equation}
where
\[\mathbf{E}=\left[ \begin{array}{cc}
-(\omega_L+\omega_H) & -\omega_L \\
\omega_H & 0 \end{array}\right],\text{   }
\mathbf{F}=\left[ \begin{array}{c}
\omega_L \\
0 \end{array}\right], \text{ and  }
\mathbf{G}=\left[ \begin{array}{cc}
1 & 0 \end{array}\right].
\]
Here $\mathbf{u}_i$ is a 2$\times$1 vector describing the state of the digital filter at node $i$, and $x_i(t)$ is the observed variable, the normalized voltage of the electrical input to the Mach-Zehnder modulator. The nodes are coupled by the adjacency matrix $\mathbf{A}=A_{ij}$; for the case of identical global coupling considered here, $A_{ij}=0$ for $i=j$ and $A_{ij}=1$ otherwise. Thus for our four node network, the number of incoming links $n_{in}=3$ for all nodes. We emphasize that the coupling is not Laplacian in general, and only becomes Laplacian in the limit $\tau_c=\tau_f$. In this work, we fix the round-trip gain $\beta=3.8$ and the feedback time delay $\tau_f=1.4$ ms and vary the global coupling strength $\varepsilon$ and coupling time delay $\tau_c$.  The filter high and low pass frequencies are given by $\omega_H/2\pi=100$ Hz and $\omega_L/2\pi=2.5$ kHz, respectively, and $\mathbf{E}$, $\mathbf{F}$, and $\mathbf{G}$ are matrices that describe the filter.

For each trial of the experiment, the nodes are initialized from noise by recording the random electrical signal into the digital signal processing (DSP) board for 50 ms. Then feedback is turned on without coupling for 500 ms in order for transients to die out. At the end of this period, the coupling is turned on for 1450 ms. We use the last 400 ms of recording to determine which synchronous state is exhibited by the network.

\subsection{Observation of partial synchronization}
We name each synchronous state by the number of nodes in each cluster. By ``clusters,'' we mean groups of synchronized nodes. Thus for a network of four nodes, the five possible states of synchrony are (Fig. 2): (a) the globally synchronized state, (b) the doublet-doublet state, (c) the triplet-singlet state, (d) the doublet-singlet-singlet (DSS) state, and the desynchronized state (not shown).  We refer to doublet-doublet and triplet-singlet as ``cluster states,'' and DSS as a ``chimera state.'' 

We observe all possible synchronous states in the experiment, as shown in Fig. 2(e-h), including a chimera state that persists for many delay times and appears to be stable. For realizations from different initial conditions, nodes appear in different clusters, confirming that the partially synchronous patterns are not a result of parameter mismatch between the oscillators. As far as we are aware, this is the first time a chimera state has been experimentally observed in such a small network. In fact, this is the minimal network of globally coupled oscillators that can support a chimera state \cite{ashwin2015weak,bohm2015amplitude}. 

\begin{figure}[h!]
\includegraphics[width=0.45\textwidth]{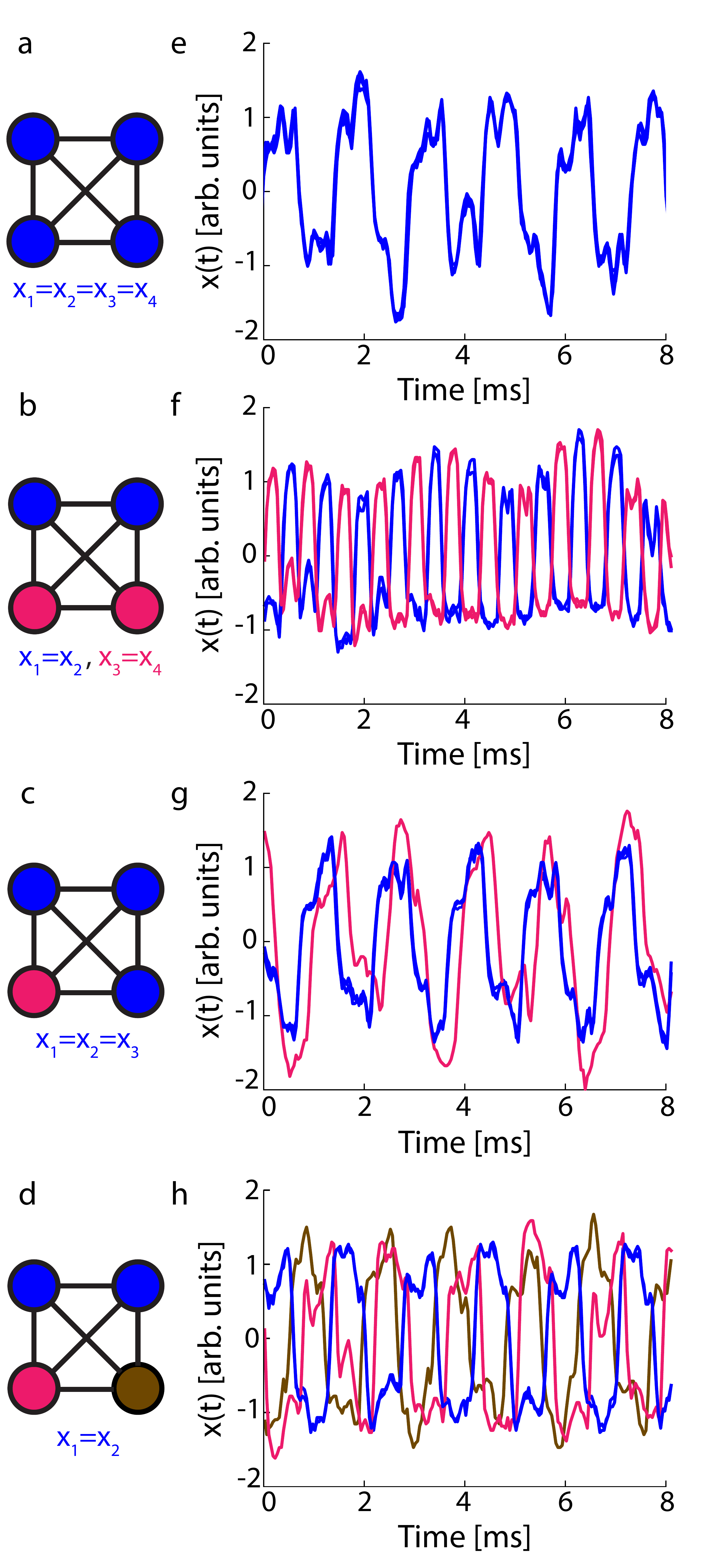}
\caption{\label{fig:timeseries}Experimentally observed synchronous states. (a)-(d) Illustration of all synchronous states for a globally coupled network of four nodes. Nodes of the same color are part of the same cluster. (e)-(h) Typical experimental time series for (e) global synchrony, (f) doublet-doublet synchrony, (g) triplet-singlet synchrony, (h) doublet-singlet-singlet synchrony (chimera). This global state was observed with $\varepsilon=0.40$ and $\tau_c=1.8$ ms, the doublet-doublet and triplet-singlet states with $\varepsilon=0.45$ and and $\tau_c=1.8$ ms, and the chimera state with $\varepsilon=0.40$ and $\tau_c=2.3$ ms.  Numerical simulation of equations (\ref{eq:ueq}) and (\ref{xeq}) gives similar time series.} 
\end{figure}
\section{Synchronization patterns and network symmetries}
 Recent studies have shown that symmetries in a network's topology determine the patterns of synchronization that the network can display \cite{pecora2014cluster,sorrentino2015complete}. We briefly review the main results. The symmetries of the adjacency matrix form a mathematical group. The nodes which are permuted among one another by these symmetries (i.e., the orbits of the group) make up the full symmetry clusters. The orbits of the subgroups of the symmetry group give the partial (subgroup) symmetry clusters that can emerge via symmetry-breaking. 

We now show that these techniques can be applied to chimeras in globally coupled networks of identical oscillators. For a globally coupled network of $N$ nodes, the nodes are indistinguishable, so the group of permutation symmetries of the adjacency matrix is the symmetric group $S_N$ (the group of all the permutations that can be performed on $N$ nodes). Since any node can be permuted with any other node, the orbit of the symmetric group is all of the nodes, and the maximal symmetry case is global synchrony. To understand the allowed partial symmetry cases, the subgroups of the symmetry group must be considered. The orbits of the subgroups of $S_N$ are such that \textit{any} partition of the $N$ oscillators is allowed to exist. In particular, a chimera state (that is, a state of one large synchronized cluster of $N_s$ oscillators and $N-N_s$ singlet ``clusters'') is permitted by the equations of motion. Whether these chimera states are possible to observe is determined by the linear stability analysis.

\subsection{Stability analysis}
To calculate the stability of the allowed synchronization patterns we use group theoretical techniques recently developed for the analysis of cluster synchronization \cite{pecora2014cluster,sorrentino2015complete}. The technique transforms the adjacency matrix to a block diagonalized form through a coordinate transformation that preserves the structure of the partially synchronous state.  The advantage of this technique is that the transformation matrix $\mathbf{T}$ also transforms the variational equations, decoupling the motion along the synchronization manifold from the directions transverse to it. As a result, the stability of the partially synchronous state can be calculated by considering only the (lower dimensional) equations for the transverse directions. We refer to the original coordinate system as the node coordinates and the transformed coordinate system as the irreducible representation (IRR) coordinates.

As an example, we describe the steps for calculating the stability of (DSS) chimera states. The stability for the other partially synchronous states was determined in a similar manner. 

It is straightforward to determine the equations of motion of the partially synchronous state under consideration (in this case, the (DSS) chimera state). The variational equations are determined by considering the time evolution of a small perturbation  $\Delta{\mathbf{u}}$ to the synchronous state and are given by

\begin{equation}
\frac{d}{dt}\Delta{\mathbf{u}}_{i}(t)=\mathbf{E}\Delta\mathbf{u}_{i}(t)+\mathbf{F}\beta\sin(2x_{i}(t)+2\phi_0) \Delta x_{i}(t)
\end{equation}
\begin{equation}
\Delta x_i(t) = \mathbf{G}\big[(1-\varepsilon)\Delta\mathbf{u}_i(t-\tau_f)
+ \frac{\varepsilon}{3}\sum_{j}{A_{ij}\Delta\mathbf{u}_j(t-\tau_c)}\big]
\end{equation}
where the $x_i(t)$ is the behavior of node $i$ in the desired partially synchronous state and we have used the fact that the network contains four globally coupled nodes.

\begin{figure*}
\includegraphics[width=0.8\textwidth]{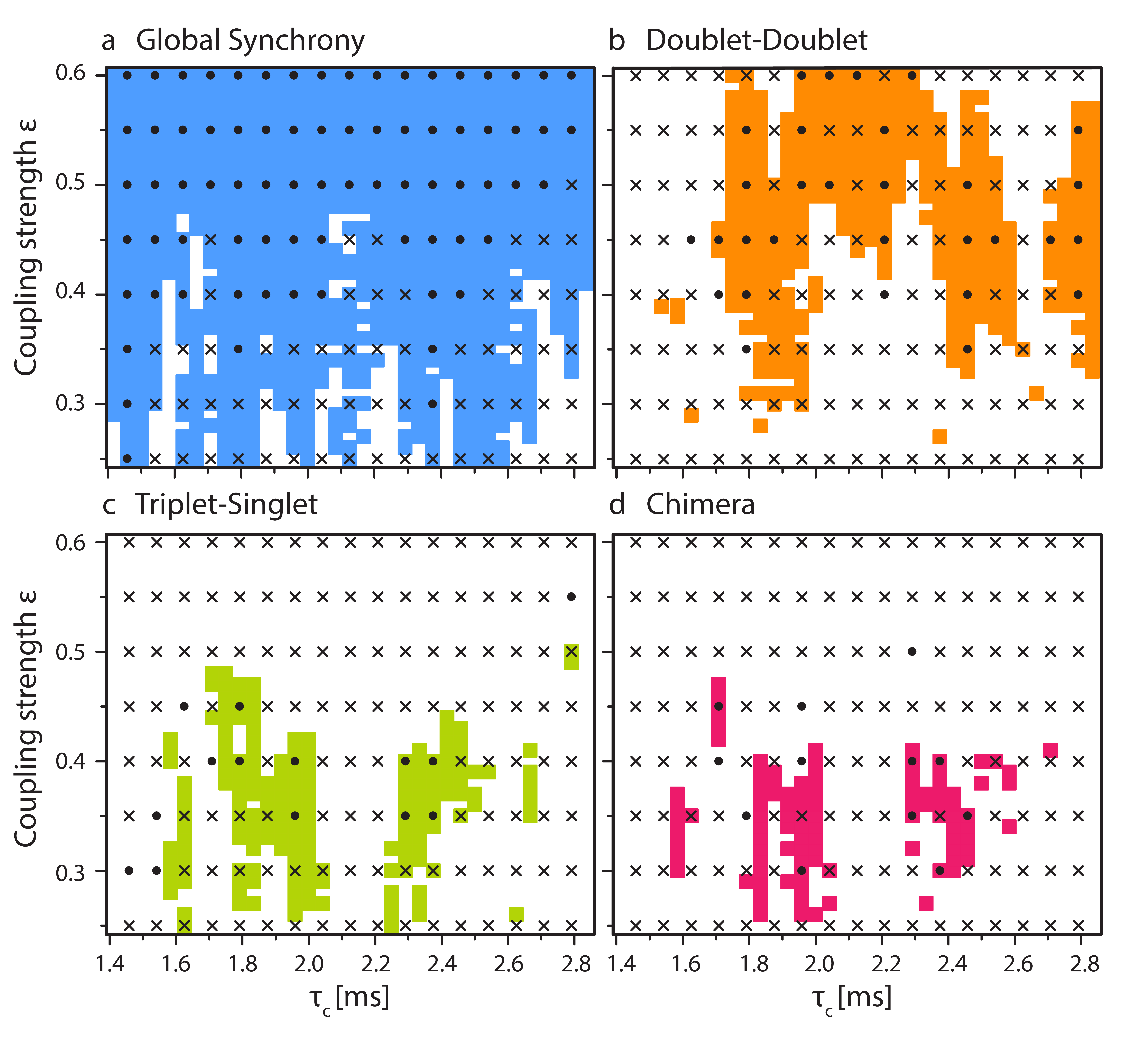}
\caption{\label{fig:Stability Results} Comparison of experimental results and stability calculations. (a)-(d) Region of stability for different synchronous states. The shaded regions are stable; that is, the LLE of the variational equations is negative. The markers represent experimental results. Dots indicate that the state has been observed in experiments; crosses indicate that the state was not observed in experiments. At least 20 trials from different random initial conditions were performed for each experimental data point. For both experiments and simulations, the round-trip gain $\beta=3.8$ and the feedback time delay $\tau_f=1.4$ ms.} 
\end{figure*}
In order to decouple the variational equations corresponding to perturbations transverse to the synchronization manifold from those corresponding to perturbations along the synchronization manifold, we now transform to the IRR coordinate system. As discussed in the appendix, the transformation matrix for the (DSS) chimera state is
\begin{equation}
\mathbf{T=}\begin{bmatrix}
\frac{1}{\sqrt{2}} & \frac{1}{\sqrt{2}} & 0 & 0 \\
0 & 0 & 1 & 0 \\
0 & 0 & 0 & 1 \\
\frac{1}{\sqrt{2}} & -\frac{1}{\sqrt{2}} & 0 & 0
\end{bmatrix},
\end{equation}
where the three upper rows correspond to the synchronization manifold and the bottom row corresponds to the direction transverse to the synchronization manifold. In order to determine the stability of the partially synchronous state, we need to consider only perturbations along directions transverse to the synchronization manifold. If we define IRR basis vectors $\Delta \mathbf{v}_i(t) \equiv T_{ij}\Delta\mathbf{u}_i(t)$, then $\Delta\mathbf{v}_4(t)$ is the only IRR basis vector corresponding to perturbations transverse to the synchronization manifold. Thus in the following we consider only $\Delta\mathbf{v}_\perp(t)\equiv \Delta\mathbf{v}_4(t)$. Applying $\mathbf{T}$ to transform the variational equations to the IRR coordinate system, we obtain

\begin{equation}\label{eq:perp1}
\frac{d}{dt} \Delta\mathbf{v}_\perp(t)=\mathbf{E}\Delta\mathbf{v}_\perp(t)+\mathbf{F}\beta\sin(2x_{s}(t)+2\phi_0) \Delta x_\perp(t)
\end{equation}
\begin{equation}\label{eq:perp2}
\Delta x_\perp(t) = \mathbf{G}\big[(1-\varepsilon)\Delta\mathbf{v}_\perp(t-\tau_f)
+ \frac{\varepsilon}{3}\sum_j{B_{ij}}\Delta\mathbf{v}_j(t-\tau_c)\big]
\end{equation}

where $x_s(t)$ is the behavior of one node in the synchronized cluster and

\begin{equation}\label{eq:Bmatrix}
\mathbf{B} = \mathbf{TAT}^{-1}=
\begin{bmatrix}
1 & \sqrt{2} & \sqrt{2} & 0 \\
\sqrt{2} & 0 & 1 & 0 \\
\sqrt{2} & 1 & 0 & 0 \\
0 & 0 & 0 & -1
\end{bmatrix}
\end{equation}
is the adjacency matrix transformed to the IRR coordinates. Explicitly performing the sum in equation (\ref{eq:perp2}), we obtain
\begin{equation}\label{eq:perp3}
\Delta x_\perp(t) = \mathbf{G}\big[(1-\varepsilon)\Delta\mathbf{v}_\perp(t-\tau_f)
- \frac{\varepsilon}{3}\Delta\mathbf{v}_\perp(t-\tau_c)\big].
\end{equation}

To determine the stability, we calculated the largest Lyapunov exponent (LLE) of equations (\ref{eq:perp1}) and (\ref{eq:perp3}), which indicates how infinitesimal perturbations transverse to the synchronization manifold grow or decay in time. If the LLE  is negative, perturbations decay exponentially to zero, indicating that the state is stable. For our calculations we used discrete-time versions of the equations presented above which are more suitable for the experimental conditions. A similar procedure was followed to obtain the stability for the other synchronous states. 

In Fig. 3, we compare the results of experiments and stability calculations in the parameter space of coupling strength ($\varepsilon$) and coupling delay ($\tau_c$) for all the partially synchronous states that the system displays. Experiments were performed by selecting regularly spaced points in the parameter space. A minimum of 20 experimental trials from different random initial conditions were performed for each point in parameter space. In principle, one can experimentally observe any state in the parameter space that theoretically shows stable solutions; however, in practice it can be difficult to observe states with small basins of attraction. As discussed in ref. \cite{menck2013basin}, the size of the basin of attraction is of great practical interest, and in future work we hope to investigate whether symmetries in the network topology can help to shed light on basin stability. For all four synchronous states, agreement between experimentally observed states and their calculated stability is quite good. However, the slight disagreement can be attributed to the finite number of random initial conditions that were used for the calculations and experiments.  

\begin{figure}
\centering
\includegraphics[width=0.45\textwidth]{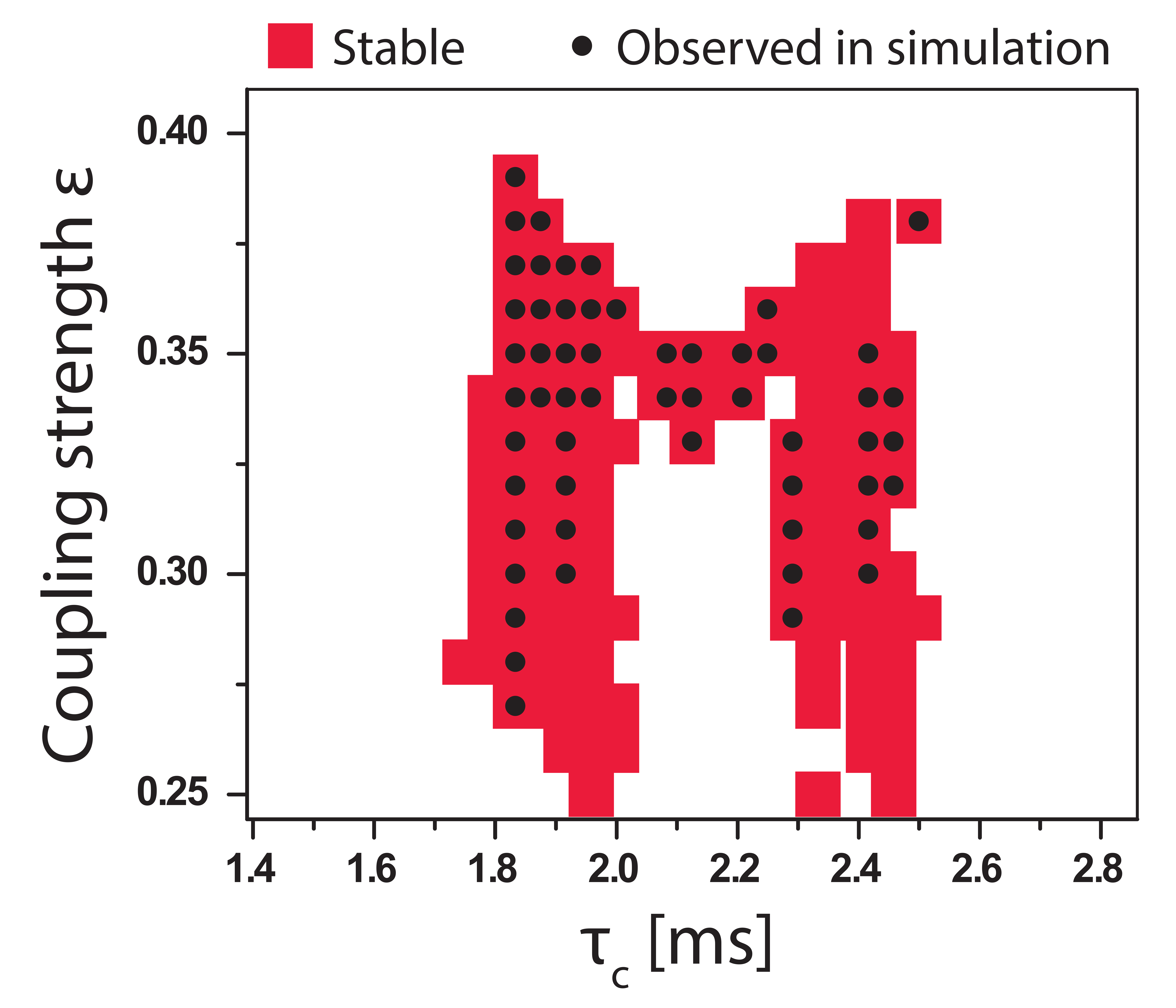}
\caption{\label{fig:QS5}Comparison of stability calculations of a chimera state (one cluster of 5 and 5 singlet clusters) with direct simulation of equations (\ref{eq:ueq}) and (\ref{xeq}) for a 10 node, globally coupled network. Direct simulations consisted of 100 trials with different initial conditions for each point in $\varepsilon-\tau_c$ parameter space. For both experiments and simulations, the round-trip gain $\beta=3.8$ and the feedback time delay $\tau_f=1.4$ ms.}
\end{figure}

The procedure for stability calculations described above can in principle be used to determine the stability of partially synchronous states (clusters and chimeras) in networks of any size. While our experiment is restricted to four nodes, we have performed the same type of stability analysis for a chimera state in a 10 node network consisting of one cluster of 5 and 5 singlet clusters, and found that it agrees with direct simulations of equations (\ref{eq:ueq}) and (\ref{xeq}), as shown in Fig. \ref{fig:QS5}.

\section{Chimera states and multistability}
Recently, the existence of chimeras has been theoretically associated with multistability in the system \cite{bohm2015amplitude}. Our observations support this idea.

\begin{figure}
\includegraphics[width=0.45\textwidth]{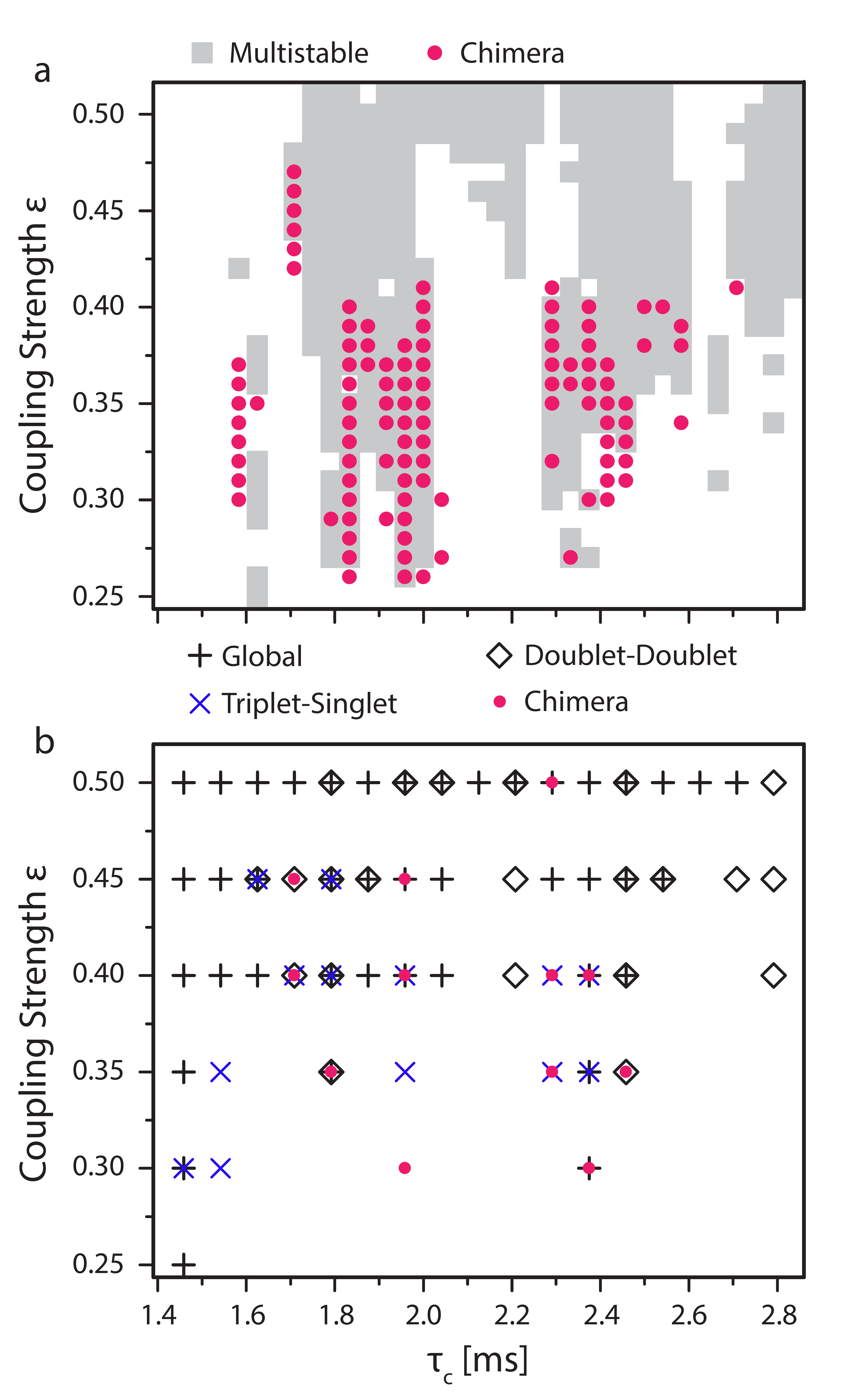}
\caption{\label{fig:Multistability} Relationship between chimeras and multistability in the four node, globally coupled network. (a) Stability calculations show that (DSS) chimeras tend to be stable in the regions of parameter space where other synchronization patterns coexist. Here multistable means that at least two of the globally synchronized, doublet-doublet, and triplet-singlet states are stable. (b) Experiments also show that chimeras tend to show up in regions of multistability. For both experiments and simulations, the round-trip gain $\beta=3.8$ and the feedback time delay $\tau_f=1.4$ ms. }
\end{figure}

In Fig. \ref{fig:Multistability} we show a direct connection between multistability and chimeras in our network of four oscillators. From the stability calculations we identified the regions where at least two of the globally synchronized, doublet-doublet, and triplet-singlet states are stable. In Fig. \ref{fig:Multistability}a such regions are marked as multistable. Calculated stable chimera solutions coincide well with these multistable regions.  In experiments, we also observe this multistability for the parameter values that exhibit chimera states, as shown in Fig. \ref{fig:Multistability}b.

In addition to the multistability between different partially synchronous patterns, we also observe multistability within a single pattern. For example, while the dynamics of the globally synchronized state in Fig. 2e appear chaotic, there are other globally synchronized states which appear to be quasiperiodic. We do not distinguish between different dynamical behaviors of the same partially synchronous state; for example, in Fig. 3a if any of the possible dynamical realizations of global synchrony is stable, we consider the globally synchronized pattern to be stable.

As discussed previously, partial synchronization patterns (doublet-doublet, triplet-singlet, and (DSS) chimera states in this case) emerge from the partial (subgroup) symmetries in the network \cite{pecora2014cluster,sorrentino2015complete}. In systems like ours this can be predicted by a detailed inspection of all the subgroup symmetries of the network by analyzing the adjacency matrix. Still, what mechanism breaks the maximal symmetry is an interesting question. For our particular system, it is the presence of two different time delays in the system introduced by the mismatch between the coupling delay and the feedback delay. When these two time delays exactly match the Laplacian coupling case), we observe only global synchrony which is also supported by previous work on these opto-electronic oscillator networks \cite{williams2013experimental}.

We understand that the dependence of the resulting synchronous state on initial conditions leads to interesting fragmented regions of stability. As the initial conditions change, the high multistability (both within a single synchronous pattern as well as between different synchronous patterns) of parameter space allows different stable states (depending upon their basins of attraction) including global synchrony and complete desynchrony. Thus, rather than observing any smooth boundary between synchronized and desynchronized regions we observe fragmented regions of stability, as  can be seen in Figs. 3 and \ref{fig:Multistability}. 

Hence, the multistability or the possibility of various partially synchronous solutions in the system can be seen as a requirement for chimeras in any system, but the physical mechanism that generates such multistability can be different for different systems. It is well-known that time-delay in the coupling can induce multistability between synchronous states (e.g., the review in ref. \cite{rodrigues2015kuramoto}). This is the case in our system, while in the laser system described in ref. \cite{bohm2015amplitude}, amplitude-phase coupling induces the multistability necessary for the appearance of chimera states.

\section{Discussion}

Our network of four globally coupled oscillators is fundamentally important in the context of chimeras. This is a small system without any breaking of symmetry in the coupling topology, yet we experimentally observe chimera states starting from random initial conditions. Our system violates all the conditions previously believed to be necessary for the formation of chimera states: it is a small network, it is initialized from random initial conditions, and it is globally coupled. Importantly, our stability calculations show that the observed chimeras are not long transients but stable physical states that persist in experiments.

The mechanism that allows the partially synchronous states to form in our system is a general phenomenon called isolated desynchronization in which some clusters separate out from the synchronized state without destroying the synchrony completely \cite{pecora2014cluster}. This is possible due to the partial (subgroup) symmetries of the network. The subgroup structure guarantees that all the nodes in one cluster receive the same effective coupling signal from nodes in other clusters. Hence even if one cluster is desynchronized, the others can remain in identical synchrony. For example, in the case of our DSS chimera, each of the oscillators in the doublet cluster receive the same total signal from the two desynchronized singlets, allowing them to remain synchronized even though the two singlet oscillators behave incoherently. The idea that the chimeras and clusters in our system arise from the same mechanism of isolated desynchronization and that their stability can be calculated in the same manner highlights the close relationship between chimera and cluster states as partial synchronization patterns. 
We emphasize that the analysis we have presented here is not restricted to globally coupled networks of oscillators. The group theoretical analysis and mechanism of isolated desynchronization extend to any network with cluster states or chimera states in which the coherent population is identically synchronized, such as those found in the non-locally coupled systems in refs. \cite{tinsley2012chimera,martens2013chimera,abrams2008solvable} and the star network in ref. \cite{sinha2015chimera}. 

While in the simulations and stability analysis we consider identical oscillators with identical coupling, some heterogeneity and noise are inevitable in experiments. Despite the small heterogeneities in our experiment, we still observe persistent chimera and cluster states, in agreement with the simulations and stability calculations. Determining the amount of heterogeneity for which the group theoretical analysis and stability calculations remain valid is an important open question.

We have observed all possible partially synchronous states, including a chimera state, in our experimental network of four globally coupled chaotic opto-electronic oscillators. We used group theoretical methods recently developed for cluster synchrony to calculate the linear stability of these states, and found excellent agreement with our experiments. These methods are quite general in that they extend to large networks and can be used to analyze the stability of any chimera state in which the coherent oscillators are identically synchronized, suggesting that such chimeras and cluster states are closely related. Our results indicate that multistability of different synchronous patterns seems to be important for the existence of stable chimera states and can be determined by analyzing the symmetries of a given network topology; however, the mechanism that generates the multistablility can be different in different systems. For our case we identify it to be the breaking of the symmetry present in Laplacian coupling by having two different time delays in the network.

\section*{Acknowledgements}
We thank F. Sorrentino, L.M. Pecora, and the authors of ref. \cite{bohm2015amplitude} (F. B\"ohm, A. Zakharova, E. Sch\"oll, and K. L\"udge)  for helpful discussions. We gratefully acknowledge ONR for supporting this work through Grant No. N000141410443.

\section*{Appendix: Calculation of the IRR transformation matrix}
In order to transform the variational equations to the IRR coordinates, the IRR transformation matrix $\mathbf{T}$ must be obtained. In general, the method of computing $\mathbf{T}$ involves finding the IRRs for the symmetry (sub)group of a given synchronous state and is nontrivial. While $\mathbf{T}$ can be computed by the method in ref. \cite{pecora2014cluster}, for the case of the DSS chimera $\mathbf{T}$ can be determined by inspection.

First, we note that for the DSS chimera, the synchronization manifold is three-dimensional: one dimension for the synchronized doublet cluster, and one dimension for each of the two singlet clusters. These can be represented by
\begin{equation}
\mathbf{v}_1=\begin{bmatrix}
\frac{1}{\sqrt{2}} \\ \frac{1}{\sqrt{2}} \\ 0 \\ 0
\end{bmatrix},
\mathbf{v}_2=\begin{bmatrix}
0 \\ 0 \\ 1 \\ 0
\end{bmatrix},
\mathbf{v}_3=\begin{bmatrix}
0 \\ 0 \\ 0 \\ 1
\end{bmatrix}
\end{equation}.

There are four nodes so the node space is four dimensional. Thus, there can be only one transverse direction: 
\begin{equation}
\mathbf{v}_4=\begin{bmatrix}
\frac{1}{\sqrt{2}} \\ -\frac{1}{\sqrt{2}} \\ 0 \\ 0
\end{bmatrix}
\end{equation}.

Then $\mathbf{T}$ can be constructed by stacking these normalized row basis vectors. If we stack them such that the vectors corresponding to the synchronization manifold are on top of the vectors corresponding to transverse directions, when we use $\mathbf{T}$ to block diagonalize the adjacency matrix the upper block will correspond to the perturbations along the synchronization manifold and the lower block will correspond to transverse perturbations. For the DSS chimera case, we find
\begin{equation}
\mathbf{T=}\begin{bmatrix}
\frac{1}{\sqrt{2}} & \frac{1}{\sqrt{2}} & 0 & 0 \\
0 & 0 & 1 & 0 \\
0 & 0 & 0 & 1 \\
\frac{1}{\sqrt{2}} & -\frac{1}{\sqrt{2}} & 0 & 0
\end{bmatrix}.
\end{equation}

\bibliography{bib}{}
\end{document}